\documentclass[aps,prx,twocolumn,longbibliography,nofootinbib]{revtex4-1}

\usepackage{amsmath,amssymb,bm}
\usepackage{graphicx}
\usepackage[colorlinks=true,linkcolor=blue,citecolor=blue,urlcolor=blue]{hyperref}
\usepackage{libertine}
\usepackage{xcolor}
\definecolor{langgreen}{rgb}{0.0,0.50,0.0}

\newcommand{\e}{\mathrm{e}}
\newcommand{\ii}{\mathrm{i}}
\newcommand{\dd}{\mathrm{d}}
\newcommand{\Tc}{T_{c}}
\newcommand{\DF}{\Delta F}
\newcommand{\vF}{v_{F}}
\newcommand{\pF}{p_{F}}
\newcommand{\aR}{\alpha_{R}}
\newcommand{\Gt}{\Gamma_{t}}
\newcommand{\xiGL}{\xi_{\rm GL}}
\newcommand{\epsB}{\epsilon_{B}}
\newcommand{\sfa}{\mathsf{a}}
\newcommand{\avg}[1]{\langle #1\rangle}
\newcommand{\Imop}{\mathrm{Im}}
\newcommand{\Reop}{\mathrm{Re}}

\begin{document}

\title{Magnetochiral anisotropy of thermally activated phase slips \\ in noncentrosymmetric superconductors}

\author{Alex Levchenko}
\affiliation{Department of Physics, University of Wisconsin--Madison, Madison, Wisconsin 53706, USA}

\date{November 28, 2026}

\begin{abstract}
Resistive tails of narrow superconducting wires below the critical temperature are governed by phase slips, which are thermally activated near the transition and quantum at lower temperatures. We develop the theory of thermally activated phase slips in wires without inversion symmetry, using as a microscopic model a diffusive wire formed from a two-dimensional electron gas with Rashba spin-orbit coupling in an in-plane magnetic field. The phase-slip resistance acquires magnetochiral anisotropy: at fixed current it changes upon reversal of the field, and at fixed field upon reversal of the current, while the linear-response resistance remains even in the field, as required by Onsager reciprocity. The nonreciprocity originates from an asymmetry of the activation barrier that is odd in both current and field. This asymmetry is generated by the cubic gradient invariant and the momentum-odd quartic vertex, together with even gradient terms that acquire odd parts in the helical ground state. The same combination of couplings determines the superconducting diode effect, so the barrier asymmetry is fixed by the diode efficiency of the same wire through a universal numerical factor. The kinetic Lifshitz invariant, which makes the relaxation and noise of the order parameter nonreciprocal, affects only the rate prefactor and is parametrically subleading. Because the asymmetry enters the activation exponent, the magnetochiral signal exceeds the diode asymmetry by a factor of the order of the barrier height divided by temperature. It appears as a field-antisymmetric splitting of the exponential slopes of the differential-resistance traces commonly used to identify thermally activated phase slips.
\end{abstract}

\maketitle

\section{Introduction}
\label{sec:intro}

Superconductivity in quasi-one-dimensional wires is qualitatively shaped by fluctuations of the order parameter. As anticipated by Little \cite{Little1967}, a thin wire below its critical temperature $\Tc$ retains a small but finite resistance because thermal fluctuations occasionally drive the order parameter amplitude to zero at some point along the wire, allowing the phase to slip by $2\pi$ and, through the Josephson relation, generating a voltage. The quantitative theory of thermally activated phase slips (TAPS) was constructed by Langer and Ambegaokar \cite{Langer1967}, who identified the saddle-point configuration of the Ginzburg--Landau (GL) free energy and the associated activation barrier, and by McCumber and Halperin \cite{McCumber1970}, who computed the fluctuation prefactor of the phase-slip rate from time-dependent GL (TDGL) dynamics. The resulting Langer--Ambegaokar--McCumber--Halperin (LAMH) resistance was confirmed in classic experiments on superconducting whiskers \cite{Lukens1970,Newbower1972} and, later, in lithographically defined and molecular-templated nanowires \cite{Giordano1988,Bezryadin2000,Lau2001,Zgirski2005,Tian2005,Altomare2006,Rogachev2005,Bollinger2006,Sahu2009}, where LAMH fits describe the resistive transition over many orders of magnitude in resistance and remain applicable in strong magnetic fields \cite{Rogachev2005}. At lower temperatures and in the thinnest wires, quantum phase slips take over \cite{Giordano1988,Zaikin1997,Golubev2001,Lau2001}. For reviews, see Refs.~\cite{Arutyunov2008,Bezryadin2008}.

A parallel development concerns superconductors without an inversion center \cite{BauerSigrist,Smidman2017}. When time-reversal symmetry is additionally broken by a Zeeman or exchange field, the interplay of parity and time-reversal breaking produces a family of magnetoelectric phenomena rooted in the Lifshitz invariants of the GL free energy: gradient terms of odd order that are forbidden in centrosymmetric materials. Their best-known consequence is the helical superconducting state, in which the condensate develops a spontaneous phase gradient set by the field \cite{Edelstein1989,MineevSamokhin1994,Agterberg2003,BarzykinGorkov2002,Samokhin2004,Kaur2005,DimitrovaFeigelman2007}. In recent years the most visible manifestation has been the superconducting diode effect (SDE), the inequivalence of the depairing critical currents for opposite current directions, observed in a growing set of noncentrosymmetric films, heterostructures, and Josephson devices \cite{Ando2020,Baumgartner2022} and studied theoretically from several complementary perspectives \cite{DaidoYanase2022,YuanFu2022,HeLaw2022,IlicBergeret2022,Ilic2024,Hasan2024,Hasan2025,Bankier2025}; for reviews see Refs.~\cite{Nadeem2023,ShafferLevchenko2025}. The SDE is a thermodynamic effect, determined by the condensation energy of uniform current-carrying states. The focus of the field is now shifting toward nonequilibrium and nonlinear transport phenomena of noncentrosymmetric superconductors, including electrical magnetochiral anisotropy (MCA) \cite{Rikken2001,RikkenWyder2005,Wakatsuki2017,WakatsukiNagaosa2018,Hoshino2018,Itahashi2020,Miranda2026a,Miranda2026b}, as well as photogalvanic effects, second-harmonic generation, nonlinear and photovoltaic Hall responses \cite{Parafilo2022,Tanaka2024,Dong2025,Miranda2026c}; for broad overviews of nonreciprocal transport and optics in quantum materials see Refs.~\cite{TokuraNagaosa2018,NagaosaYanase2024}.

Transport coefficients differ from thermodynamic ones in an essential way. In the GL description the SDE is controlled by the Lifshitz invariants of the static free energy, whereas a dc transport coefficient also depends on the kinetic sector of the theory, namely the dissipative dynamics and the noise of the order parameter. In a recent work \cite{Liu2026} the generalized TDGL theory of a disordered Rashba superconductor was derived from the Keldysh nonlinear sigma model, and it was found that the relaxation rate of a fluctuation with pair momentum $q$, and hence by the fluctuation-dissipation theorem the Langevin noise power, acquires a term odd in $q$ and odd in the field. This kinetic Lifshitz invariant does not affect equilibrium thermodynamics, but it contributes to nonlinear fluctuation transport at the same order as the thermodynamic invariants; above $\Tc$, for instance, it carries half of the fluctuation-induced MCA \cite{Liu2026}. A similar separation into thermodynamic and kinetic contributions is expected for other nonreciprocal transport coefficients. Beyond the fluctuation regime, the theory of nonreciprocal kinetics in superconductors is largely limited to quasiparticle mechanisms of the Debye type, in which nonreciprocity is amplified by the long inelastic relaxation time \cite{Smith2021,LiuSmith2024}.

In this paper we develop the theory of thermally activated phase slips in a noncentrosymmetric superconducting wire. The problem connects the two lines of research outlined above, since phase slips are the elementary resistive fluctuations of a thin wire and the Lifshitz invariants describe its broken inversion and time-reversal symmetries. The thermodynamic sector of the generalized GL functional controls the activation exponent: the phase-slip barrier acquires a correction that is odd in both the bias current and the field, so that forward and backward phase slips at the same current are activated over different barriers for opposite field orientations. The kinetic sector, including the kinetic Lifshitz invariant, controls the fluctuation prefactor, and we show that its nonreciprocal corrections are parametrically subleading, smaller than the exponent asymmetry by the ratio of temperature to barrier height. The barrier asymmetry is determined by the same combination of Lifshitz invariants that fixes the diode efficiency $\eta$ of the same wire, including the known cancellations among the cubic gradient invariant, the momentum-odd quartic vertex, and the even gradient corrections promoted to odd terms by the helical shift \cite{Hasan2024,Hasan2025,Liu2026}. The two observables are related without adjustable parameters: the anomalous phase of the phase-slip saddle equals $\sqrt6\,\eta$. Because this asymmetry enters the activation exponent, the relative magnetochiral change of the resistance exceeds the diode efficiency by a factor of order the barrier height over temperature, roughly an order of magnitude in the regime where TAPS are observed. The effect appears as a field-antisymmetric splitting of the exponential slopes of differential-resistance traces of the type measured in Ref.~\cite{Rogachev2005}, and vanishes when the in-plane field is aligned with the wire, which gives a characteristic angular dependence.

The paper is organized as follows. Section~\ref{sec:model} introduces the microscopic model and the generalized TDGL theory, defines the thermodynamic and kinetic Lifshitz invariants, and summarizes the physics of the helical state and of the diode effect that follows from the same functional. Section~\ref{sec:taps} contains the central calculation: the phase-slip saddle point of the noncentrosymmetric wire, the exact cancellation of the leading Lifshitz invariant at fixed current, the first-order theory of the subleading invariants, the resulting barrier asymmetry, and the analysis of the fluctuation prefactor. Section~\ref{sec:results} presents the transport results: the nonreciprocal current-voltage characteristic, the low-field expansion of the magnetoresistance, experimental signatures, quantitative estimates, and the range of validity. Section~\ref{sec:discussion} summarizes the results, compares them with prior work on phase slips beyond the standard LAMH framework, and outlines extensions. Technical material is collected in four appendices: the closed-form geometry of the phase-slip saddle (Appendix~\ref{app:saddle}), the resolution of gradient-ordering ambiguities (Appendix~\ref{app:ordering}),  
the fluctuation determinant and the prefactor controversy (Appendix~\ref{app:prefactor}), and the microscopic values of all couplings across the spin-relaxation crossover (Appendix~\ref{app:coeffs}). We set $\hbar=k_{B}=1$ except in final transport formulas.

\section{Model and generalized TDGL theory}
\label{sec:model}

\subsection{Geometry and symmetry considerations}
\label{sec:geometry}

We consider a narrow diffusive superconducting wire patterned from a two-dimensional electron gas with Rashba spin-orbit coupling of strength $\aR$, elastic scattering time $\tau$, diffusion constant $D=v^2_F\tau/2$, and an in-plane Zeeman field $h$ (from an applied field $B$, $h=\tfrac12 g\mu_{B}B$, or from exchange with an adjacent magnet). The wire axis is $\hat x$, the Rashba (polar) axis $\hat n_{R}$ is normal to the electron plane, and the field lies in the plane. The in-plane polar vector $\hat n \parallel \hat n_{R}\times\bm h$ is odd under time reversal, and all nonreciprocal couplings of the theory are directed along it. We orient the field perpendicular to the wire, so that $\hat n$ is aligned with the wire axis and every effect discussed below is maximal; rotating the in-plane field toward the wire axis multiplies all odd couplings by $\sin\theta_{B}$, where $\theta_{B}$ is the angle between $\bm B$ and the wire. The wire is thinner than the coherence length and the penetration depth, so the order parameter $\Delta(x)$ depends only on the coordinate along the wire and orbital screening currents are negligible; $s$ denotes the cross-sectional area and $\nu$ the normal density of states.

A Lifshitz invariant is a free-energy term of odd order in gradients. For a single complex order parameter the real part of $\Delta^{*}\partial_{x}\Delta$ is a total derivative, so the lowest-order invariant is $\Imop(\Delta^{*}\partial_{x}\Delta)$, which changes sign under time reversal. Its coefficient must therefore be odd under time reversal, and it requires both broken inversion symmetry, provided by the Rashba axis, and broken time-reversal symmetry, provided by $\bm h$. Every odd coupling below is therefore odd in $B$. The same couplings control the diode effect, which is the origin of the connection established in this paper.

\subsection{Free energy with Lifshitz invariants}
\label{sec:freeenergy}

Our starting point is the generalized GL functional of the disordered Rashba superconductor derived in Ref.~\cite{Liu2026} from the Keldysh nonlinear sigma model \cite{Levchenko2007,Virtanen2022}. Reduced to the wire geometry it reads
\begin{align}
F[\Delta]={}&s\nu\!\int\!\dd x\,\Big[
\epsilon_{h}|\Delta|^{2}
+\xiGL^{2}\,|\partial_{x}\Delta|^{2}
-\Lambda\,\Imop(\Delta^{*}\partial_{x}\Delta)
\nonumber\\
&+\Xi\,\Imop(\partial_{x}\Delta^{*}\,\partial_{x}^{2}\Delta)
-\alpha_{4}|\partial_{x}^{2}\Delta|^{2}
+\frac{b_{0}}{2}|\Delta|^{4}
\nonumber\\
&+\frac{b_{0}\mathfrak{b}_1}{2}\,|\Delta|^{2}\,\Imop(\Delta^{*}\partial_{x}\Delta)
+\ldots\Big],
\label{eq:F}
\end{align}
with the standard dirty-limit even coefficients
\begin{equation}
\epsilon_{h}=\frac{T-\Tc}{T}+\frac{7\zeta(3)h^{2}}{4\pi^{2}T^{2}},
\quad
\xiGL^{2}=\frac{\pi D}{8T},
\quad
b_{0}=\frac{7\zeta(3)}{8\pi^{2}T^{2}} .
\label{eq:evencoeffs}
\end{equation}
The three terms odd in gradients are the Lifshitz invariants. For a uniform helix $\Delta=|\Delta|\e^{\ii qx}$ their densities are $-\Lambda q|\Delta|^{2}$, $\Xi q^{3}|\Delta|^{2}$, and $(b_{0}\mathfrak{b}_1/2)q|\Delta|^{4}$: the linear invariant $\Lambda$, the cubic gradient invariant $\Xi$, and the momentum-odd part of the quartic vertex $\mathfrak{b}_1$. The ellipsis stands for the even vertex gradient correction, which at plane-wave level renders the quartic coefficient momentum dependent,
\begin{equation}
b(q)=b_{0}\big(1+\mathfrak{b}_1\,q-\mathfrak{b}_{2}\,q^{2}\big),
\label{eq:bq}
\end{equation}
and for higher even gradients; the $q^{4}$ coefficient $-\alpha_{4}$ is retained explicitly in Eq.~\eqref{eq:F} because, as first emphasized in the diode context \cite{Hasan2024,Liu2026}, even terms of this order contribute to nonreciprocal observables once the expansion is referenced to the helical ground state.

All couplings follow from a single closed-form kernel controlled by the ratio of the Dyakonov--Perel (DP) spin-relaxation rate $\Gt=2(\aR\pF)^{2}\tau$ to temperature \cite{Liu2026}. Physically, the Lifshitz invariants arise from a second-order process in which a singlet pair fluctuation converts into a triplet via the Zeeman field, propagates as a triplet subject to DP relaxation, and converts back via the spin-orbit field-strength vertex; the triplet lifetime therefore controls the size of every invariant. Two limits will be quoted throughout: the weak-relaxation regime $\Gt\ll4\pi T$, where all odd couplings are suppressed by $(\aR\pF\tau)^{2}$, and the relaxation-dominated regime $\Gt\gg4\pi T$, where they saturate. The Zeeman term of $\epsilon_{h}$ in Eq.~\eqref{eq:evencoeffs} is written in its weak-relaxation form; in the relaxation-dominated regime it is reduced to $\pi h^{2}/2T\Gt$ \cite{Liu2026}. The full crossover forms are collected in Appendix~\ref{app:coeffs}. A convenient summary is the helical wave vector
\begin{equation}
q_{0}=\frac{\Lambda}{2\xiGL^{2}}
=\frac{4\aR h}{\vF^{2}}\;\mathfrak{g}\Big(\frac{\Gt}{4\pi T}\Big),
\label{eq:q0}
\end{equation}
where $\mathfrak{g}(g)\simeq14\zeta(3)g/\pi^{2}$ for $g\ll1$ and $\mathfrak{g}\to1$ for $g\gg1$: in the DP-dominated regime the helical modulation saturates at a universal, disorder-independent value. In practice the crossover scale $\aR\pF\sim\sqrt{T/\tau}$ is parametrically smaller than the boundary $\aR\pF\tau\sim1$, so even weakly spin-orbit-coupled wires are often already in the universal regime near $\Tc$ \cite{Liu2026}.

One property of the microscopic coefficients is important below. The full momentum dependence of the quartic vertex is a convolution over three independent pair momenta; evaluating the uniform kernel at a shifted cooperon pole underestimates $\mathfrak{b}_1$ by the factor $8/3$ and $\mathfrak{b}_{2}$ by $4/3$ \cite{Liu2026,Hasan2025}. Since the diode effect, and as we show below the phase-slip asymmetry, are controlled by near-cancellations among these couplings, the ladder-complete vertex values of Appendix~\ref{app:coeffs} are essential; with the truncated vertex even the sign of the net effect is incorrect.

\subsection{Stochastic dynamics and the kinetic Lifshitz invariant}
\label{sec:dynamics}

Because the theory of Ref.~\cite{Liu2026} is formulated on the Keldysh contour, it determines the dissipative dynamics and the noise of the order parameter in addition to the free energy. The relaxation kernel of a pair fluctuation with momentum $q$ acquires a momentum-odd part,
\begin{equation}
\Imop L_{R}^{-1}(q,\omega)=\frac{\pi\omega}{8T}\,\big[1+\rho\,q\big],
\label{eq:kinetic}
\end{equation}
where the kinetic coefficient $\rho\propto\aR h$ is odd in the field (its value across the DP crossover is given in Appendix~\ref{app:coeffs}). The corresponding stochastic TDGL equation for the wire is
\begin{align}
\frac{\pi}{8T}\big(1-\ii\rho\,\partial_{x}\big)_{\rm sym}\,\partial_{t}\Delta
&=-\frac{1}{s\nu}\frac{\delta F[\Delta]}{\delta\Delta^{*}}+\xi_{\Delta}(x,t),
\label{eq:TDGL}
\end{align}
with the Hermitian symmetrization of the momentum-odd friction operator understood, and with Langevin noise whose power carries the same nonreciprocal factor,
\begin{equation}
\avg{\xi_{\Delta}\xi_{\Delta}^{*}}(q,\omega)
=\frac{\pi\omega}{8T\nu s}\coth\Big(\frac{\omega}{2T}\Big)\big[1+\rho\,q\big].
\label{eq:noise}
\end{equation}
Fluctuations with pair momenta parallel and antiparallel to $\hat n$ therefore decay at different rates. Such a momentum-odd kinetic coefficient is admissible because $h$ is odd under time reversal; the Onsager relation $\Imop L_{R}^{-1}(q,\omega,\bm h)=\Imop L_{R}^{-1}(-q,\omega,-\bm h)$ permits a term proportional to $(\hat n_{R}\times\bm h)\cdot q$ while forbidding any $q$-odd friction that is even in the field. It requires $\rho(\bm h)=-\rho(-\bm h)$, which the microscopic result $\rho\propto\aR h$ satisfies. By the fluctuation-dissipation theorem, friction and noise acquire the same factor $1+\rho q$, Eqs.~\eqref{eq:kinetic} and \eqref{eq:noise}. As a result, in the classical regime $\omega\ll T$ relevant near $\Tc$, the stationary distribution of Eq.~\eqref{eq:TDGL} is the Gibbs distribution $\propto\e^{-F[\Delta]/T}$, independent of $\rho$. This exact property is central to the analysis below. The kinetic invariant thus changes relaxation rates but not equilibrium probabilities. It cannot enter activation exponents, and its effect on phase slips is limited to the rate prefactor, a statement we quantify in Sec.~\ref{sec:prefactor}.

\subsection{Helical state, depairing currents, and the diode effect}
\label{sec:sde}

Before turning to phase slips we summarize the physics of uniform current-carrying states encoded in Eq.~\eqref{eq:F}, both to fix notation and because the diode effect is the thermodynamic counterpart of the phase-slip asymmetry obtained below. For plane waves $\Delta=|\Delta|\e^{\ii qx}$ the condensation problem is governed by
\begin{align}
\frac{F}{s\nu}&=a(q)|\Delta|^{2}+\frac{b(q)}{2}|\Delta|^{4},
\nonumber\\
a(q)&=\epsilon_{h}-\Lambda q+\xiGL^{2}q^{2}+\Xi q^{3}-\alpha_{4}q^{4},
\label{eq:aq}
\end{align}
with $b(q)$ of Eq.~\eqref{eq:bq}. For $a(q)<0$ the amplitude adjusts to $|\Delta|^{2}=-a/b$, and the supercurrent follows from differentiating the free energy at fixed amplitude,
\begin{equation}
j(q)=2e\nu\Big[a'(q)|\Delta|^{2}+\tfrac12 b'(q)|\Delta|^{4}\Big].
\label{eq:jq}
\end{equation}
The second term, which originates from the momentum-odd vertex, contributes to the diode asymmetry at the same order as the first. Reducing to a momentum-independent quartic coefficient by the substitution $\sfa(q)=a(q)\sqrt{b_{0}/b(q)}$, so that $F_{\rm cond}/s\nu=-\sfa^{2}(q)/2b_{0}$, and shifting to the helical minimum $q=q_{0}+k$, one finds that all odd couplings combine into a single effective cubic coefficient \cite{Hasan2024,Liu2026},
\begin{align}
\sfa(k)&=\epsB-\tfrac12\epsB\big(\mathfrak{b}_1-2\mathfrak{b}_{2}q_{0}\big)k+\xiGL^{2}k^{2}
\nonumber\\
&\quad+a_{3}k^{3}-\alpha_{4}k^{4}+\dots,
\nonumber\\
\epsB&=\epsilon_{h}-\frac{\Lambda^{2}}{4\xiGL^{2}}<0,
\label{eq:ashift}
\end{align}
\begin{equation}
a_{3}=\Xi-\frac{\xiGL^{2}\mathfrak{b}_1}{2}
-\frac{2\Lambda\alpha_{4}}{\xiGL^{2}}+\frac{\Lambda\mathfrak{b}_{2}}{2}.
\label{eq:a3}
\end{equation}
Here $k$ denotes the deviation of the pair momentum from the helical minimum; throughout the paper $k$ has this meaning, and $\epsB$ plays the role of the reduced distance to the field-renormalized transition, combining Zeeman pair breaking with the condensation-energy gain of the helical state. The term linear in $k$ in Eq.~\eqref{eq:ashift} is of first order in the odd couplings. It displaces the zero-current state by $\delta k=\epsB(\mathfrak{b}_1-2\mathfrak{b}_{2}q_{0})/4\xiGL^{2}$ without changing the extremal values of the current to this order, and therefore does not enter the diode efficiency. The linear invariant drops out of the quadratic form, since it is absorbed by the helical shift. It nevertheless contributes to $a_{3}$ through the products $\Lambda\alpha_{4}$ and $\Lambda\mathfrak{b}_{2}$, because the shift converts even gradient terms into odd ones. The critical currents are the extrema of Eq.~\eqref{eq:jq} on the two branches, $k_{c}^{\pm}\simeq\pm\sqrt{|\epsB|/3\xiGL^{2}}$, and to first order in the odd couplings the diode efficiency is
\begin{equation}
\eta\equiv\frac{j_{c}^{+}-|j_{c}^{-}|}{j_{c}^{+}+|j_{c}^{-}|}
=\frac{\sqrt{|\epsB|}\;a_{3}}{\sqrt3\,\xiGL^{3}} .
\label{eq:eta}
\end{equation}
In the weak-DP regime the four contributions to $2a_{3}/\xiGL^{2}$, in units of $(\aR\pF\tau)^{2}\aR h/T^{2}$, are
\begin{equation}
1-\frac{124\,\zeta(5)}{7\zeta(3)\pi^{2}}-\frac{392\,\zeta^{2}(3)}{\pi^{6}}+\frac{2}{3}
\simeq-0.471 ,
\label{eq:bracket}
\end{equation}
so that the cubic invariant (the first entry) is overcompensated by the quartic vertex and the $q^{4}$ gradient term, and the polarity of the effect is opposite to what the cubic invariant alone would give \cite{Liu2026}. These cancellations are characteristic of the Rashba model. We show below that the phase-slip barrier asymmetry is governed by the same combination \eqref{eq:a3}, with the same relative weights of all four terms.

\section{Nonreciprocal thermally activated phase slips}
\label{sec:taps}

\subsection{Phase-slip saddle point and activation barriers}
\label{sec:saddle}

We first set up the phase-slip problem in the frame of the helical ground state, where it takes the conventional LAMH form, and treat the odd couplings as perturbations. Writing $\Delta(x)=\e^{\ii q_{0}x}\psi(x)$ and measuring lengths in units of the coherence length $\xi(T)=\xiGL/\sqrt{|\epsB|}$, the order parameter in units of the equilibrium amplitude $\Delta_{B}=(|\epsB|/b_{0})^{1/2}$, and energies in units of
\begin{equation}
E_{\xi}=\frac{s\nu\,\epsB^{2}}{b_{0}}\,\xi(T)
=\frac{H_{c}^{2}}{4\pi}\,s\,\xi(T),
\label{eq:Exi}
\end{equation}
the even sector of Eq.~\eqref{eq:F} reduces to the standard dimensionless GL functional for $u\equiv \psi/\Delta_{B}=f\e^{\ii\phi}$,
\begin{equation}
F[u]=\int\dd x\,\Big[|u'|^{2}-|u|^{2}+\tfrac12|u|^{4}+\tfrac12\Big],
\label{eq:GLdimless}
\end{equation}
whose stationary configurations obey $u''+u-|u|^{2}u=0$ with the conserved dimensionless supercurrent $j=\Imop(u^{*}u')=f^{2}\phi'$. The higher even gradients ($\alpha_{4}$, $\mathfrak{b}_{2}$ proper) modify Eq.~\eqref{eq:GLdimless} only at relative order $|\epsB|$ and are consistently neglected in the saddle geometry; their essential role is the odd contributions they generate after the helical shift, retained in full below. The physical current is related to $j$ by the identity
\begin{equation}
\frac{\hbar I}{2e}=2E_{\xi}\,j ,
\label{eq:current-identity}
\end{equation}
which relates saddle-point integrals to transport quantities.

\begin{figure*}[t]
\centering
\includegraphics[width=\textwidth]{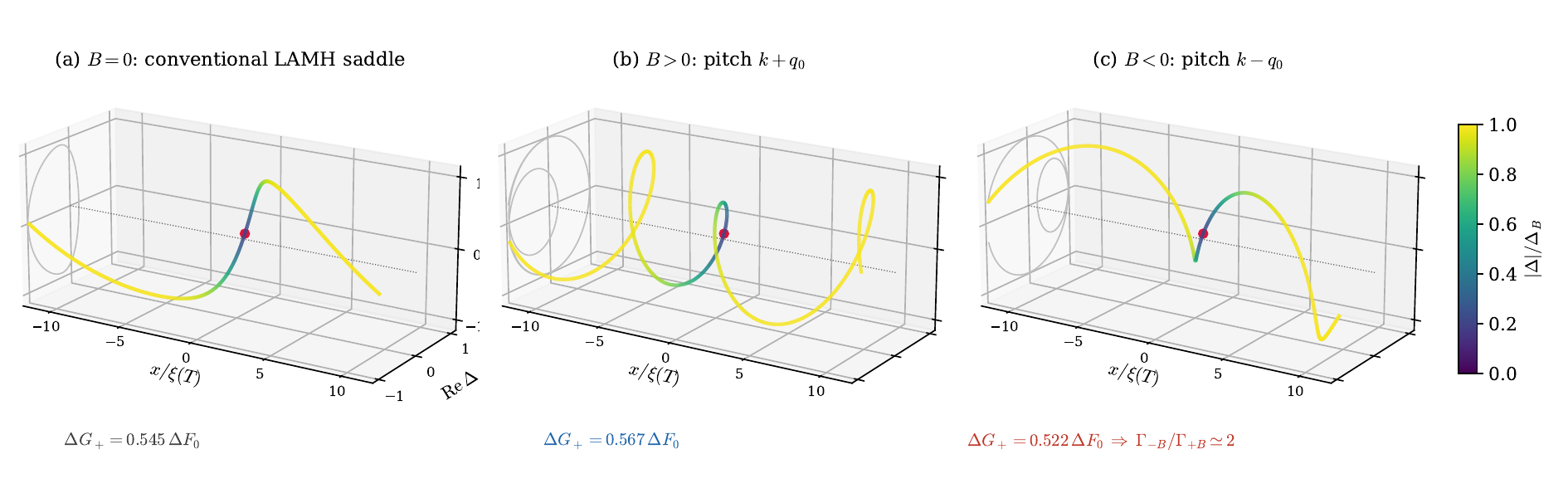}
\caption{Phase-slip saddle point of the noncentrosymmetric wire, plotted from the closed-form solution, Eqs.~\eqref{eq:LAsaddle} and \eqref{eq:phase}: the trajectory of $\Delta(x)$ along the wire (color encodes $|\Delta|/\Delta_{B}$; gray curve shows the projection onto the complex-$\Delta$ plane; the red dot marks the slip core at $x=0$, where $|\Delta|_{\min}=\sqrt2\,k\,\Delta_{B}\simeq0.21\,\Delta_{B}$; the dotted line is the $\Delta=0$ axis). Parameters: $k=0.15$ (fixed by the bias current, $I\simeq0.38\,I_{c}$, identical in all panels), helical wave vector $q_{0}\xi=\pm0.5$, anomalous phase $\chi=\pm0.15$ corresponding to a diode efficiency $|\eta|=\chi/\sqrt6\simeq6\%$, and $\DF_{0}/k_{B}T=15$. (a) $B=0$: the conventional Langer--Ambegaokar saddle with background pitch $k$. (b) $B>0$: the same core superimposed on the helical ground state with total pitch $k+q_{0}$. (c) $B<0$: pitch $k-q_{0}$; for $q_{0}>k$ the background winding reverses chirality while carrying the same current. The slip core and the phase twist $\delta\phi(k)\simeq0.86\pi$ across it are orientation independent at leading order; the Lifshitz couplings instead shift the forward barriers to $\Delta G_{+}=0.545,\,0.567,\,0.522\times\DF_{0}$ for panels (a), (b), (c), so the slip rate at this current differs by a factor $\Gamma_{-B}/\Gamma_{+B}\simeq2$ between the two field orientations. The barrier and phase offset obtained by numerical integration of the plotted configuration agree with the closed forms of Appendix~\ref{app:saddle}.}
\label{fig:helix}
\end{figure*}

Uniform solutions are helices $u=\sqrt{1-k^{2}}\,\e^{\ii kx}$ with $j(k)=k(1-k^{2})$ and depairing at $k_{c}=1/\sqrt3$, $j_{c}=2/3\sqrt3$; recall that $k$ measures the phase gradient relative to the helical minimum. On a wire biased below the critical current these states are metastable: decay of the supercurrent requires the order parameter amplitude to be suppressed somewhere along the wire so that the phase can unwind. The lowest saddle point mediating this decay was found by Langer and Ambegaokar \cite{Langer1967}. Integrating the amplitude equation once and imposing uniform boundary conditions gives, for the squared amplitude $f^{2}(x)$,
\begin{align}
f^{2}(x)&=(1-k^{2})-(1-3k^{2})\,{\rm sech}^{2}(\beta x),
\nonumber\\
\beta&=\sqrt{\frac{1-3k^{2}}{2}},
\qquad
f^{2}(0)=2k^{2}:
\label{eq:LAsaddle}
\end{align}
a localized dip of width $\sim\xi/\beta$ superimposed on the current-carrying background. The phase profile integrates in closed form (Appendix~\ref{app:saddle}),
\begin{equation}
\phi(x)=k\,x+\arctan\Big[\frac{\sqrt{1-3k^{2}}}{\sqrt2\,k}\,\tanh(\beta x)\Big],
\label{eq:phase}
\end{equation}
so the saddle of the noncentrosymmetric wire in the laboratory frame, $\Delta_{s}(x)=\e^{\ii q_{0}x}f(x)\e^{\ii\phi(x)}$, is known explicitly. Its geometry is shown in Fig.~\ref{fig:helix}: the trajectory of $\Delta(x)$ in the complex plane is a helix pinched at the slip core, where the amplitude dips to $|\Delta|_{\min}=\sqrt2\,k\,\Delta_{B}$ without vanishing; only in the zero-current limit does the saddle degenerate into a real kink with a true node. Reversing the field reverses $q_{0}$ and hence the winding of the background helix at fixed current, a direct manifestation of the helical state; the slip core itself is unchanged at this order.

Two saddle integrals are needed below: the condensate deficit of the dip and the extra phase it winds relative to the uniform state at the same current,
\begin{align}
\int\dd x\,\big(f^{2}-f_{\infty}^{2}\big)&=-2\sqrt2\,\sqrt{1-3k^{2}},
\label{eq:deficit}\\
\delta\phi(k)&=2\arctan\frac{\sqrt{2(1-3k^{2})}}{2k}\in(0,\pi].
\label{eq:deltaphi}
\end{align}
The free-energy cost of the saddle relative to the metastable state at the same current has the closed form
\begin{equation}
\DF(k)=\DF_{0}\,\sqrt{1-3k^{2}},
\qquad
\DF_{0}=\frac{8\sqrt2}{3}\,\frac{H_{c}^{2}}{8\pi}\,s\,\xi(T),
\label{eq:DFk}
\end{equation}
which in dirty-limit microscopic units reads $\DF_{0}=\frac{16\pi^{2}}{21\zeta(3)}s\nu\sqrt{\pi D}\,(T_{c}-T)^{3/2}$, with $T_c$ understood as the field-renormalized transition temperature implied by $\epsB$.

At fixed bias current the relevant energy is the Gibbs free energy $G=F-(\hbar I/2e)\,\Delta\varphi$, which accounts for the work done by the current source as the total phase difference $\Delta\varphi$ changes. Because each phase slip changes $\Delta\varphi$ by $2\pi$, the work per slip,
\begin{equation}
\delta W=\frac{\pi\hbar I}{e},
\label{eq:work}
\end{equation}
is fixed by the $2\pi$ periodicity of the phase and is therefore independent of the Lifshitz couplings. The forward and backward Gibbs barriers can be organized as
\begin{equation}
\Delta G_{\pm}=\Delta\bar F\mp\frac{\pi\hbar I}{2e},
\qquad
\Delta\bar F=\DF(k)-\frac{\hbar I}{2e}\big[\delta\phi(k)-\pi\big].
\label{eq:Gpm}
\end{equation}
Forward and backward slips pass through the same saddle, so their prefactors cancel in the ratio $\Gamma_{+}/\Gamma_{-}=\exp(\delta W/k_{B}T)$, as required by detailed balance. Together with the Josephson relation $V=(\pi\hbar/e)(\Gamma_{+}-\Gamma_{-})$ this yields the familiar structure
\begin{equation}
V=\frac{\pi\hbar}{e}\,2\,\Omega\,
\e^{-\Delta\bar F/k_{B}T}\sinh\Big(\frac{\pi\hbar I}{2ek_{B}T}\Big),
\label{eq:VI}
\end{equation}
with a common attempt frequency $\Omega$. In a centrosymmetric wire the offset $\delta\phi(k)-\pi=-2\sqrt2\,k+\mathcal{O}(k^{3})$ is odd in $k$, so the correction to $\Delta\bar F$ is even in the current: time-reversal symmetry forbids any term odd in $I$ alone. In the next two subsections we show that broken inversion and time-reversal symmetries generate such a term, with a coefficient odd in $B$.

For the prefactor we use the Kramers--Langer escape rate of the stochastic dynamics \eqref{eq:TDGL}. Evaluating the fluctuation determinant around the saddle in closed form (Appendix~\ref{app:prefactor}) gives, at small currents,
\begin{equation}
\Omega=\frac{\sqrt3}{2\pi^{3/2}}\,\frac{L}{\xi(T)}\,\frac{1}{\tau_{\rm GL}}
\Big(\frac{\DF_{0}}{k_{B}T}\Big)^{1/2},
\quad
\tau_{\rm GL}=\frac{\pi\hbar}{8k_{B}(\Tc-T)},
\label{eq:MH}
\end{equation}
which coincides with the McCumber--Halperin result \cite{McCumber1970}: the rate is proportional to the wire length, contains the factor $(\DF_{0}/k_{B}T)^{1/2}$ from the translational zero mode of the slip, and is set by the GL relaxation time. Authors of Ref. \cite{Golubev2008} argued that the attempt frequency is of order $\Tc$ rather than $\tau_{\rm GL}^{-1}$. In Appendix~\ref{app:prefactor} we show that the static fluctuation determinants of the two calculations coincide and that the disagreement is confined to the dynamical conversion factor, and we explain why Eq.~\eqref{eq:MH} is the internally consistent choice for the microscopically derived overdamped dynamics \eqref{eq:TDGL} in the regime considered here. None of the nonreciprocal observables constructed below depend on this choice.

\subsection{Exact cancellation of the linear Lifshitz invariant at fixed current}
\label{sec:cancellation}

By analogy with the anomalous phase of $\varphi_{0}$ Josephson junctions \cite{Buzdin2008}, one might expect the nonreciprocity of phase slips to originate from the leading Lifshitz invariant $\Lambda$. This is not the case. In the shifted frame $\Delta=\e^{\ii q_{0}x}\psi$ the quadratic sector of Eq.~\eqref{eq:F} maps exactly onto that of a conventional superconductor,
\begin{align}
\epsilon_{h}|\Delta|^{2}+\xiGL^{2}|\partial_{x}\Delta|^{2}
-\Lambda\Imop(\Delta^{*}\partial_{x}\Delta)
\nonumber\\
=\epsB|\psi|^{2}+\xiGL^{2}|\partial_{x}\psi|^{2},
\label{eq:gauge}
\end{align}
the total supercurrent of this sector is the conventional current of $\psi$, and the constant background winding $q_{0}L$ cancels from every barrier difference in the Gibbs functional. At fixed current, therefore, the linear invariant has no effect on the phase-slip problem beyond the even-in-$B$ renormalization $\epsilon_{h}\to\epsB$. All odd-in-$B$ phase-slip physics is carried by the subleading momentum-odd couplings.

Two remarks ensure that the analysis below captures the cancellations characteristic of the Rashba model. First, the reference helix may be chosen as the minimizer of the quadratic sector alone, $q_{0}=\Lambda/2\xiGL^{2}$, without loss of generality: any first-order redefinition $q_{0}\to q_{0}+\delta q_{0}$ changes the functional by a term proportional to $\Imop(\psi^{*}\partial_{x}\psi)$, which on any stationary configuration equals the conserved current density $j$, constant along the wire and identical for the saddle and the metastable state at the same bias. Such terms integrate to $jL$ in both configurations and cancel from every barrier difference. For the same reason, the term linear in $k$ in the plane-wave expansion \eqref{eq:ashift}, which amounts to such a redefinition, does not contribute to the fixed-current barrier. Second, the plane-wave coefficients $a(q)$ and $b(q)$ do not fix the real-space operators uniquely; different Hermitian orderings differ by amplitude-gradient terms that vanish on plane waves but not on the saddle. This ambiguity, however, is confined to the even sector at the order considered: after the shift, the part of any even operator that is linear in $q_{0}$ reduces to a unique cross term. First-order results in the odd couplings are therefore free of gradient-ordering ambiguities; an explicit demonstration is given in Appendix~\ref{app:ordering}.

\subsection{First-order theory of the odd couplings}
\label{sec:oddtheory}

Working in the shifted frame and keeping every term of first order in the odd couplings, the perturbation to the conventional functional \eqref{eq:GLdimless} consists of four terms: the cubic invariant $\Xi$; the odd quartic vertex $\mathfrak{b}_1$; the odd part of the $\alpha_{4}$ term generated by the shift, whose operator content is identical to that of the cubic invariant with coefficient $-4\alpha_{4}q_{0}$; and the odd part of the $\mathfrak{b}_{2}$ vertex correction, identical in form to the odd quartic vertex, which replaces $\mathfrak{b}_1$ by $\mathfrak{b}_1-2\mathfrak{b}_{2}q_{0}$. Only two effective couplings can therefore enter,
\begin{equation}
\Xi_{\rm eff}=\Xi-4\alpha_{4}q_{0},
\qquad
\mathfrak{b}_{\rm eff}=\mathfrak{b}_1-2\mathfrak{b}_{2}q_{0},
\label{eq:effcouplings}
\end{equation}
multiplying the two odd operator structures
\begin{equation}
\mathcal{O}_{3}=\Imop(\psi'^{*}\psi''),
\qquad
\mathcal{O}_{1}=|\psi|^{2}\,\Imop(\psi^{*}\psi') .
\label{eq:operators}
\end{equation}
In the dimensionless units of Sec.~\ref{sec:saddle} the perturbation reads
\begin{align}
\delta F_{\rm odd}[u]=E_{\xi}\sqrt{|\epsB|}\int\dd x\,
\Big[&\frac{\Xi_{\rm eff}}{\xiGL^{3}}\,\Imop(u'^{*}u'')
\nonumber\\
&+\frac{\mathfrak{b}_{\rm eff}}{2\,\xiGL}\,|u|^{2}\Imop(u^{*}u')\Big].
\label{eq:dFodd}
\end{align}

The barrier correction follows from first-order perturbation theory. Because the unperturbed saddle and metastable state are both stationary points of the Gibbs functional at fixed current, the first-order shift of the barrier is given by $\delta F_{\rm odd}$ evaluated on the unperturbed configurations, with errors quadratic in the odd couplings. No re-solution of the saddle equations is required. This step fails only within a parametrically narrow window at depairing, where the saddle softens; we return to that regime in Sec.~\ref{sec:diode-link}.

Two on-shell identities simplify the evaluation. With $u=f\e^{\ii\phi}$ and $\phi'=j/f^{2}$, the odd quartic density is
\begin{equation}
|u|^{2}\Imop(u^{*}u')=j\,f^{2},
\label{eq:quarticdensity}
\end{equation}
the product of the current and the local condensate density. The odd vertex thus assigns to the moving condensate an energy that depends on the direction of motion, and the amplitude dip becomes less costly for one current direction than for the other. For the cubic operator the saddle-point equation reduces the density to
\begin{equation}
\Imop(u'^{*}u'')=j\,\big(1-f^{2}\big).
\label{eq:cubicdensity}
\end{equation}
Equation~\eqref{eq:cubicdensity} is the key step. On shell, the cubic-invariant density equals, up to a constant, minus the quartic-invariant density. Both perturbations therefore reduce to the same saddle integral, the condensate deficit \eqref{eq:deficit}, and the barrier depends only on the combination $\Xi_{\rm eff}/\xiGL^{2}-\mathfrak{b}_{\rm eff}/2$, which also determines the diode efficiency. Carrying out the integrals,
\begin{align}
\delta\Delta\bar F
&=E_{\xi}\sqrt{|\epsB|}\;j\,\Big[\frac{\Xi_{\rm eff}}{\xiGL^{3}}-\frac{\mathfrak{b}_{\rm eff}}{2\xiGL}\Big]
\times2\sqrt2\,\sqrt{1-3k^{2}}
\nonumber\\
&=\frac{\hbar I}{2e}\,\chi\,\sqrt{1-3k^{2}},
\label{eq:dFbar}
\end{align}
where Eq.~\eqref{eq:current-identity} was used in the second line and the anomalous phase $\chi$ collects the couplings. Substituting $q_{0}=\Lambda/2\xiGL^{2}$ into Eq.~\eqref{eq:effcouplings} one finds that the bracket reproduces the diode combination \eqref{eq:a3} exactly, term by term,
\begin{equation}
\chi=\sqrt2\,\sqrt{|\epsB|}\,\frac{a_{3}}{\xiGL^{3}}
=\sqrt{6}\;\eta .
\label{eq:chi}
\end{equation}
The phase-slip barrier asymmetry is controlled by the same effective cubic coefficient $a_{3}$ that determines the diode efficiency, including all four contributions of Eq.~\eqref{eq:bracket} with identical relative weights, and the relation between the two observables is parameter free. In the weak-DP regime, explicitly,
\begin{equation}
\chi\simeq-0.333\,\frac{(\aR\pF\tau)^{2}\,\aR h}{T^{2}}\,
\frac{\sqrt{|\epsB|}}{\xiGL},
\label{eq:chiweak}
\end{equation}
odd in the field through $\aR h$, vanishing as $\sqrt{|\epsB|}$ at the transition, and increasing below the transition in the same way as $\eta$.

The physical interpretation of Eq.~\eqref{eq:dFbar} is an anomalous phase offset of the phase-slip saddle. Writing $\Delta G_{\pm}=\DF(k)-\frac{\hbar I}{2e}[\delta\phi_{\rm eff}-\pi]\mp\frac{\pi\hbar I}{2e}$ with $\delta\phi_{\rm eff}=\delta\phi(k)-\chi\sqrt{1-3k^{2}}$, one sees that the wire behaves, with respect to its fluctuations, as a distributed analog of a $\varphi_{0}$ junction \cite{Buzdin2008}: the subleading Lifshitz couplings shift the phase offset of the saddle. This correction has three properties. It is odd under $I\to-I$ and under $B\to-B$ separately and even under their combination, which is the symmetry of magnetochiral anisotropy. The work per slip, Eq.~\eqref{eq:work}, is unchanged, so detailed balance and the $\sinh$ form of Eq.~\eqref{eq:VI} are preserved. Finally, it contains the same factor $\sqrt{1-3k^{2}}$ as $\DF(k)$, so at fixed current it amounts to a rescaling of the barrier; below we use the additive form \eqref{eq:dFbar}.

\subsection{Fluctuation dynamics and the prefactor}
\label{sec:prefactor}

We now estimate the corrections of the Lifshitz couplings to the rate prefactor and show that they are subleading. Since the stationary distribution is the Gibbs distribution (Sec.~\ref{sec:dynamics}), the kinetic invariant $\rho$ affects neither the activation exponent nor the ratio of forward and backward rates; it enters only the escape kinetics.

Two odd corrections to the attempt frequency arise at first order. The friction operator in Eq.~\eqref{eq:TDGL} depends on the direction of the local pair momentum, so the growth rate of the unstable mode, which is built on the background gradient $q_{0}+k/\xi$, is modified by a factor $1-\rho\,(q_{0}+c_{1}k/\xi)$ with $c_{1}$ of order unity. In addition, the static Hessians around the saddle acquire first-order corrections from the second variation of the odd operators \eqref{eq:operators}, shifting the fluctuation determinants and zero-mode norms by relative amounts of order $\chi k$. Collecting both sources,
\begin{equation}
\Omega(I,B)=\Omega_{0}(T,B^{2})\,
\big[1-\rho\,q_{0}-\big(c_{1}\partial_{B}\hat\rho+c_{2}\partial_{B}\chi\big)kB+\ldots\big],
\label{eq:Omegaodd}
\end{equation}
with $c_{1,2}=\mathcal{O}(1)$ and $\hat\rho=\rho/\xi(T)=\rho\sqrt{|\epsB|}/\xiGL$ the dimensionless kinetic coupling. The term $\rho q_{0}\propto h^{2}$ is even in the field and merely renormalizes the prefactor at order $B^{2}$. The odd terms are proportional to $kB$, as symmetry requires, and their size relative to the exponent asymmetry is
\begin{equation}
\frac{(\delta\Omega/\Omega)_{\rm odd}}{(\delta\Delta\bar F/k_{B}T)_{\rm odd}}
\sim\Big(1+\frac{|\hat\rho|}{|\chi|}\Big)\frac{k}{j}\,\frac{k_{B}T}{2E_{\xi}}\sim\frac{k_{B}T}{\DF_{0}}\ll1 ,
\label{eq:prefactor-ratio}
\end{equation}
uniformly across the LAMH window, since both odd terms scale linearly in $k$ at small current, provided $|\hat\rho/\chi|\lesssim1$. This condition holds generically but fails near the special points where $a_{3}$, and with it $\chi$ and $\eta$, changes sign as a function of disorder (Appendix~\ref{app:coeffs}). Near such a point the kinetic invariant becomes the leading source of nonreciprocity, and the relation $\chi=\sqrt6\,\eta$ no longer controls the signal. In the regime where phase slips are observable as an exponentially small resistance, $\DF_{0}\gtrsim10\,k_{B}T$, prefactor nonreciprocity is therefore a percent-level correction to the barrier effect. It becomes comparable only in the Ginzburg region $\DF_{0}\lesssim k_{B}T$, where the TAPS description crosses over to the Gaussian fluctuation regime above $\Tc$, in which the kinetic and cubic invariants are known to contribute to the magnetochiral anisotropy at the same order \cite{Liu2026}; the two descriptions are expected to connect smoothly across the transition. We also note that the attempt-frequency controversy discussed in Appendix~\ref{app:prefactor} does not affect these conclusions: in either scheme the attempt factor is even in $(I,B)$ at leading order, and the static determinants, identical in the two calculations, do not depend on $\rho$.

\subsection{Consistency with the diode effect}
\label{sec:diode-link}

The relation $\chi=\sqrt6\,\eta$ was derived at currents well below depairing, where perturbation theory in the odd couplings is controlled. An independent check is provided by the vicinity of the depairing current. Near depairing the unperturbed barrier vanishes as $\Delta G_{+}\propto(j_{c}-j)^{5/4}$, while the perturbation \eqref{eq:dFbar} vanishes only as $(j_{c}-j)^{1/4}$; expanding both and comparing with the barrier obtained from the unperturbed theory with the critical current shifted to its diode values $j_{c}^{\pm}=j_{c}(1\pm\eta)$, one finds that the two expressions coincide if $\chi=\sqrt6\,\eta$ (Appendix~\ref{app:saddle}). The barrier asymmetry computed from the small-current saddle integrals is thus the asymmetry required for the barrier to vanish at the diode-shifted critical currents. In other words, the diode effect marks the current at which the asymmetric barrier that controls thermal phase slips vanishes.

A simpler estimate, obtained by evaluating the conventional LAMH expression with $I_{c}$ replaced by $I_{c}^{\pm}(B)$, is only qualitatively correct. The ratio of the exact barrier asymmetry to this estimate is $\sqrt6\sqrt{1-3k^{2}}/\delta\phi(k)$, which tends to unity at depairing but to $\sqrt6/\pi\simeq0.78$ at small currents: the rescaling overestimates the low-current rectification by a factor $\pi/\sqrt6\simeq1.28$. At low currents the asymmetry is determined by the anomalous phase $\chi$ rather than by the shift of the critical current.

\section{Results: nonreciprocal magnetoresistance}
\label{sec:results}

\subsection{Current-voltage characteristic and the phase-slip free-energy landscape}
\label{sec:landscape}

Collecting the results of Sec.~\ref{sec:taps}, the voltage generated by thermally activated phase slips in the noncentrosymmetric wire is
\begin{align}
V(I,B)&=\frac{\pi\hbar}{e}\,2\,\Omega(I,B)\,
\e^{-\Delta\bar F(I,B)/k_{B}T}
\sinh\Big(\frac{\pi\hbar I}{2ek_{B}T}\Big),
\nonumber\\
\Delta\bar F&=\DF_{0}\sqrt{1-3k^{2}}
\nonumber\\
&\quad+\frac{\hbar I}{2e}\Big[\chi\sqrt{1-3k^{2}}-\delta\phi(k)+\pi\Big],
\label{eq:Vfull}
\end{align}
with $\chi=\sqrt6\,\eta(B,T)$ odd in the field, $k=k(I)$ the smaller root of $j(k)=k(1-k^2)$ at the given bias, and $\Omega$ given by Eq.~\eqref{eq:MH}, whose dependence on the field is even at leading order.

\begin{figure*}[t]
\centering
\includegraphics[width=\textwidth]{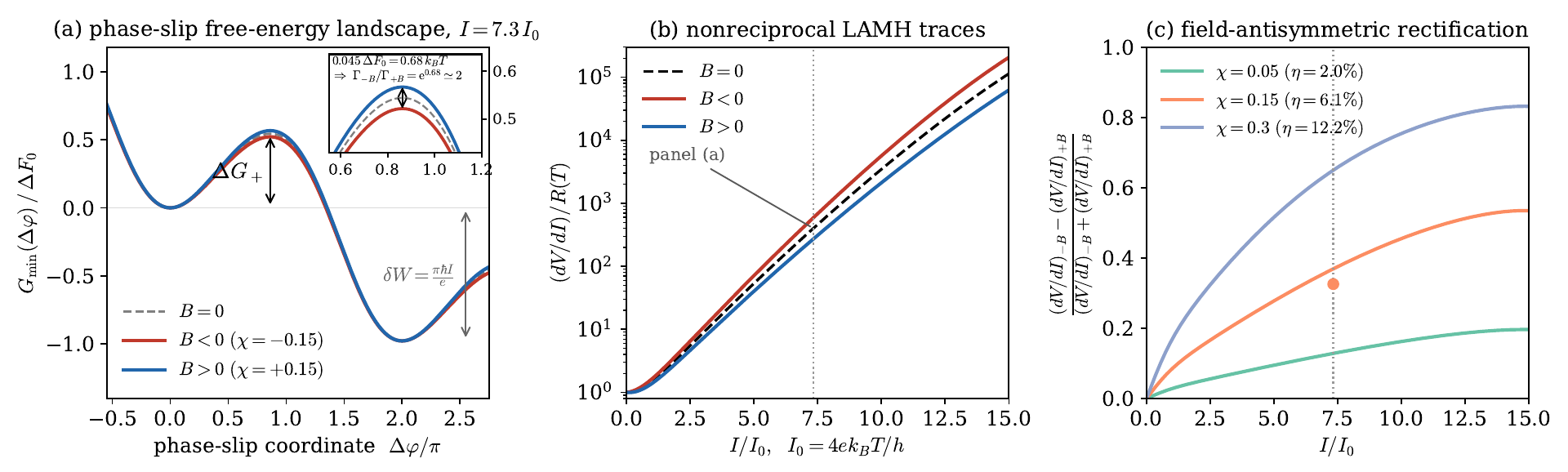}
\caption{Mechanism and observable of the nonreciprocal phase-slip magnetoresistance, with all panels sharing the parameter set of Fig.~\ref{fig:helix}: $k=0.15$, $\chi=\pm0.15$ (diode efficiency $|\eta|=\chi/\sqrt6\simeq6.1\%$), $\DF_{0}/k_{B}T=15$; energies in units of $\DF_{0}$ and currents in units of $I_{0}=4ek_{B}T/h$.
(a) Phase-slip free-energy landscape at $I=7.3\,I_{0}$ for the two field orientations (solid) and $B=0$ (dashed): a smooth interpolation through the exactly computed stationary points, with minima offset by the work per slip $\delta W=\pi\hbar I/e=0.977\,\DF_{0}$ and saddle points at phase offset $\delta\phi(k)=0.86\pi$ of height $\Delta G_{+}/\DF_{0}=0.545\mp0.023$ for $B\lessgtr0$. The curvature at the minima is schematic (it scales as $\xi/L$); extrema, locations, and tilt are exact. Inset: the saddle region; the barrier splitting $0.045\,\DF_{0}=0.68\,k_{B}T$ gives the slip-rate ratio $\Gamma_{-B}/\Gamma_{+B}=\e^{0.68}\simeq2$.
(b) Differential resistance obtained from Eq.~\eqref{eq:Vfull} with the exact $k(I)$ and a current-independent prefactor, normalized to the zero-bias resistance $R(T)$, on a logarithmic scale. At small currents the traces follow Eq.~\eqref{eq:dvdi} and the exponential slope splits by $\mp\sqrt6\,\eta/\pi$ upon field reversal; at larger currents the field-even correction to the barrier reduces the slope; all traces coincide at $I=0$, as Onsager reciprocity requires. The dotted line marks the current of panel (a), where the trace separation equals approximately the factor of two computed there. (c) Field-antisymmetrized combination of the traces of panel (b) for three values of $\chi$. This ratio is independent of $R(T)$ and of the attempt frequency, and it vanishes identically for every field-even contribution to the resistance. The dot is the rectification ratio \eqref{eq:rect} at the current of panel (a), $\tanh(0.34)\simeq0.33$; the differential ratio plotted here is somewhat larger, $0.37$, because the derivative with respect to the current also acts on the odd part of the exponent.}
\label{fig:dvdi}
\end{figure*}

Figure~\ref{fig:dvdi}(a) illustrates Eq.~\eqref{eq:Vfull} by means of the phase-slip free-energy landscape, defined as the Gibbs free energy at fixed current, minimized over all order-parameter configurations at constrained end-to-end phase. Its stationary points are exact. The minima form a ladder of physically distinct winding states, spaced by $2\pi$ and offset by the work per slip $\delta W=\pi\hbar I/e$; the barrier tops are the phase-slip saddle of Fig.~\ref{fig:helix}, located at phase offset $\delta\phi(k)$, with heights $\Delta G_{+}$ that differ for the two field orientations by $2\,(\hbar I/2e)\,\chi\sqrt{1-3k^{2}}$. The landscape resembles the tilted washboard potential of a current-biased Josephson junction, and its tilt has the same origin, the work of the current source. The analogy is, however, limited. In a junction the periodicity expresses the compactness of a single dynamical phase, whereas here it expresses the degeneracy of distinct winding states of a field theory, and the end-to-end phase is a collective coordinate rather than a dynamical degree of freedom: rates follow from the multidimensional escape theory of Sec.~\ref{sec:saddle}, not from motion of a fictitious phase particle on the plotted curve. Only the extrema, their locations, and the tilt of the landscape are universal; the curvature near the minima scales inversely with the wire length, and the interpolating curve in Fig.~\ref{fig:dvdi}(a) is schematic between stationary points.

The two field orientations shift the saddle in opposite directions, while the tilt is unchanged because the work per slip does not depend on the field. The nonreciprocity is therefore due entirely to the barrier asymmetry, which plays the role of an anomalous Josephson phase. Increasing the current steepens the tilt until the saddle merges with the minimum, and because the saddles for the two orientations have different heights, the merger occurs at different currents, $I_{c}^{\pm}=I_{c}(1\pm\eta)$.

\subsection{Low-field expansion and symmetry constraints}
\label{sec:symmetry-results}

Define the nonlinear resistance $\mathcal{R}(I,B)=V/I$. Equation~\eqref{eq:Vfull} satisfies the Onsager relation $\mathcal{R}(I,B)=\mathcal{R}(-I,-B)$ identically: under the combined reversal $k\to-k$, $\chi\to-\chi$, and every term maps onto itself. Expanding at small field and current,
\begin{equation}
\mathcal{R}(I,B)=\mathcal{R}_{0}(I,T)\Big[1-\chi'(T)\,\frac{\hbar\,IB}{2ek_{B}T}+\mathcal{R}_{2}\,B^{2}+\ldots\Big],
\label{eq:lowfield}
\end{equation}
with $\chi'=\partial_{B}\chi|_{B=0}$ and $\mathcal{R}_{0}$ the conventional LAMH resistance. Two consequences follow. First, a magnetoresistance linear in $B$ at vanishing current is forbidden: the linear-in-$B$ coefficient is necessarily odd in $I$ and vanishes in the Ohmic limit. Broken inversion and time-reversal symmetries allow only the magnetochiral term proportional to $IB$, which makes the resistance nonreciprocal, $\mathcal{R}(I,B)\neq\mathcal{R}(-I,B)$. Second, the field-even coefficient $\mathcal{R}_{2}$ contains two competing contributions of opposite sign through the barrier $\DF_{0}(\epsB)$: Zeeman pair breaking suppresses the barrier and yields positive magnetoresistance, while the helical condensation-energy gain $\Lambda^{2}/4\xiGL^{2}$ raises the transition temperature and the barrier, yielding a negative contribution. The helical gain is smaller than the Zeeman pair breaking by a factor of order $(\aR\pF\tau)^{2}(\aR/\vF)^{2}\ll1$ or less across the DP crossover, so $\mathcal{R}_{2}>0$ and the quadratic magnetoresistance is positive. These field-even effects do not contribute to the magnetochiral signal.

The nonreciprocity is characterized most directly by a ratio in which the prefactor cancels. From Eq.~\eqref{eq:Vfull},
\begin{equation}
\frac{V(I,B)+V(-I,B)}{V(I,B)-V(-I,B)}
=-\tanh\Big[\frac{\chi(B)\,\hbar I}{2ek_{B}T}\sqrt{1-3k^{2}}\Big],
\label{eq:rect}
\end{equation}
exact up to the parametrically small odd prefactor corrections of Sec.~\ref{sec:prefactor}: every field-even factor, including the attempt frequency at leading order, cancels in this ratio.  
The term $I[\delta\phi(k)-\pi]$ does not contribute, because $\delta\phi(k)-\pi$ is odd in $k$ by Eq.~\eqref{eq:deltaphi} and the identity $\arctan(1/x)=\pi/2-\arctan x$, so the product is even under $I\to-I$. At small arguments the magnitude of the rectification coefficient is $\sqrt6\,\eta\,\hbar I/2ek_{B}T$.

\subsection{Experimental signatures}
\label{sec:signatures}

The standard nonlinear characterization of TAPS is the differential resistance below the switching current, which in MoGe nanowires was shown to follow the LAMH form $dV/dI=R(T)\cosh(I/I_{0})$ with $I_{0}=4ek_{B}T/h$ over several decades, even at fields of several tesla \cite{Rogachev2005}. Our theory generalizes this fitting formula. In the LAMH window $k\ll k_{c}$, where $\sqrt{1-3k^{2}}\simeq1$ and $\delta\phi\simeq\pi$, Eq.~\eqref{eq:Vfull} gives $V=R(T)\,I_{0}\,\e^{-\chi I/\pi I_{0}}\sinh(I/I_{0})$, and hence
\begin{align}
\frac{dV}{dI}&=R(T)\,\e^{-\chi I/\pi I_{0}}
\Big[\cosh\frac{I}{I_{0}}-\frac{\chi}{\pi}\sinh\frac{I}{I_{0}}\Big],
\nonumber\\
I_{0}&=\frac{4ek_{B}T}{h}\simeq13.4\,\text{nA}\times T[\text{K}],
\label{eq:dvdi}
\end{align}
where we used $(\hbar I/2e)/k_{B}T=I/\pi I_{0}$. At $\chi=0$ this is exactly the fit of Ref.~\cite{Rogachev2005}. Equation~\eqref{eq:dvdi} neglects the field-even correction to the exponent, $(\Delta\bar F-\DF_{0})/k_{B}T\simeq\tfrac43(I/\pi I_{0})^{2}\,k_{B}T/\DF_{0}$, and therefore holds for $I\ll\pi I_{0}(\DF_{0}/k_{B}T)^{1/2}$; at larger currents the full expression \eqref{eq:Vfull} should be used, as is done in Fig.~\ref{fig:dvdi}(b). The Lifshitz couplings produce a field-antisymmetric change of the exponential slope: for $I\gg I_{0}$,
\begin{equation}
\frac{dV}{dI}\propto\exp\Big[\Big(1\mp\frac{\sqrt6\,|\eta|}{\pi}\Big)\frac{I}{I_{0}}\Big]
\quad\text{for}\quad B\gtrless0 ,
\label{eq:slopes}
\end{equation}
where the sign of $B$ is chosen such that $\chi>0$ for $B>0$, so reversing the in-plane field transverse to the wire, or equivalently reversing the current, changes the slope of $\ln(dV/dI)$ versus $I$ by $\mp\sqrt6\,\eta/\pi$, with no adjustable parameters once the diode efficiency of the same wire is known. The predicted traces are shown in Fig.~\ref{fig:dvdi}(b); the field-antisymmetrized ratio of Fig.~\ref{fig:dvdi}(c) does not depend on the zero-bias resistance or on any field-even factor that is independent of the current, and it vanishes for a field-even resistance.

Several independent tests can identify the mechanism. (i) Angular dependence: all odd couplings are proportional to $\sin\theta_{B}$, so the splitting vanishes when the in-plane field is rotated to the wire axis and reverses with the field component transverse to it. (ii) Temperature dependence: $\chi\propto\sqrt{|\epsB|}$ vanishes at the transition and increases below it in the same way as the diode efficiency, while the factor $I/\pi I_{0}$ at fixed $I/I_{c}$ grows as $\DF_{0}/k_{B}T\propto|\epsB|^{3/2}$; the measurable rectification therefore increases rapidly below $\Tc$ within the LAMH window. (iii) The parameter-free cross-check: the ratio of the measured phase-slip anomalous phase to the independently measured diode efficiency of the same wire must equal $\sqrt6$. (iv) Continuity across $\Tc$: above the transition the same symmetry channel is carried by the fluctuation magnetochiral anisotropy, with the kinetic invariant contributing on par with the cubic one \cite{Liu2026}; the crossover through the critical region connects the two regimes.

Testing the relation $\chi=\sqrt6\,\eta$ requires the diode efficiency of the depairing current, whereas the switching current $I_{\rm sw}$ measured in nanowires is reduced below $I_{c}$ by the phase slips themselves \cite{Sahu2009}. Within the present theory the asymmetry of the switching current is also fixed by the anomalous phase. If switching occurs when the phase-slip rate reaches a threshold that does not depend on the orientation of the field, the barrier $\Delta G_{+}$ at the switching current is the same for both field orientations. Using Eq.~\eqref{eq:dFbar} and the relation $\dd\Delta G_{+}/\dd I=-(\hbar/2e)\,\delta\phi(k)$, which follows from the stationarity of the Gibbs functional, we obtain the efficiency of the switching current,
\begin{equation}
\eta_{\rm sw}\equiv\frac{I_{\rm sw}^{+}-|I_{\rm sw}^{-}|}{I_{\rm sw}^{+}+|I_{\rm sw}^{-}|}
=\frac{\chi\sqrt{1-3k_{\rm sw}^{2}}}{\delta\phi(k_{\rm sw})},
\label{eq:etasw}
\end{equation}
where $k_{\rm sw}$ corresponds to $I_{\rm sw}$. The ratio $\eta_{\rm sw}/\eta=\sqrt6\sqrt{1-3k_{\rm sw}^{2}}/\delta\phi(k_{\rm sw})$ is the function introduced in Sec.~\ref{sec:diode-link}: it increases from $\sqrt6/\pi\simeq0.78$ at small currents to unity at depairing, and lies between $0.9$ and $1$ for $I_{\rm sw}\gtrsim0.5\,I_{c}$. The anomalous phase can therefore be obtained independently from the slope splitting \eqref{eq:slopes} and from the asymmetry of the switching current, and their comparison requires the depairing current only through the weak dependence of this ratio on $I_{\rm sw}/I_{c}$.

\subsection{Magnitude, parameters, and range of validity}
\label{sec:validity}

The asymmetry is enhanced because it enters the activation exponent. The nonreciprocal part of the resistance is
\begin{equation}
\frac{\mathcal{R}(I,B)-\mathcal{R}(-I,B)}{2\mathcal{R}(I,0)}
\simeq-\sqrt6\,\eta\;\frac{\hbar I}{2ek_{B}T}\,\sqrt{1-3k^{2}},
\label{eq:amplification}
\end{equation}
and the factor multiplying $\eta$ is not small in the relevant part of the LAMH regime: the combination $j\sqrt{1-3k^{2}}$ maximizes at $k\simeq0.64k_c$, i.e. $I\simeq0.83\,I_{c}$, where the odd exponent reaches $\theta_{\max}\simeq0.26\,(\DF_{0}/k_{B}T)\,\chi\simeq0.64\,(\DF_{0}/k_{B}T)\,\eta$. With $\DF_{0}/k_{B}T\sim10$--$20$, the window in which the resistive tail spans many decades, the phase-slip rectification is $(6$--$13)\times\eta$ before the hyperbolic tangent saturates: the activation exponent amplifies the bare diode asymmetry by roughly an order of magnitude. A wire with a modest measured diode efficiency of $5\%$ exhibits an asymmetry of about $30$--$55\%$ of its phase-slip resistance.

The parameters of Figs.~\ref{fig:helix} and \ref{fig:dvdi} were chosen to be mutually consistent and experimentally motivated. The reduced current $k=0.15$ corresponds to $I\simeq0.38\,I_{c}$ and, at $\DF_{0}/k_{B}T=15$, to $I=7.3\,I_{0}$; for a MoGe-like wire at $T\simeq1$--$2$~K one has $I_{0}\simeq13$--$27$~nA, so the plotted range $I\lesssim15\,I_{0}$ lies within the window explored in Ref.~\cite{Rogachev2005} before the switching current terminates the traces. The value $\chi=0.15$ corresponds to $\eta\simeq6\%$, representative of diode efficiencies reported in strongly spin-orbit-coupled platforms \cite{Ando2020,Baumgartner2022,Nadeem2023}. For the intrinsic weak spin-orbit mechanism evaluated microscopically, Eq.~\eqref{eq:chiweak} with conservative parameters ($\aR\pF\tau=0.3$, $h/T=0.5$, $T\tau=0.1$, $E_{F}\tau=50$, $|\epsB|=0.1$) gives $\chi\sim10^{-4}$: as for the diode effect itself \cite{Liu2026}, a sizable signal requires strong Rashba splitting or proximity-induced exchange. The relation $\chi=\sqrt6\,\eta$ then allows the phase-slip prediction to be calibrated against the measured diode efficiency of the same device without knowledge of the microscopic couplings.

The domain of validity of these results is bounded by four conditions. First, the GL window: $|\epsB|\ll1$ with the wire outside the Ginzburg region, $\DF_{0}\gg k_{B}T$, so that phase slips are rare and the saddle-point treatment applies; this is the same condition under which the LAMH fit itself is meaningful. Second, perturbation theory in the odd couplings requires $\chi\ll1$, that is $\eta\ll1$, satisfied in most reported diode experiments; moreover, the second-order corrections to the barrier are even in the field and cancel from the field-antisymmetrized observables. Third, the first-order treatment fails within a narrow window of the depairing current, of relative width of order $\chi$ in current (order $\sqrt{\chi}$ in the reduced momentum $k$), where the saddle softens; there the physics is instead captured by the shifted critical currents themselves, as shown in Sec.~\ref{sec:diode-link}, so the two descriptions cover the full current axis. Fourth, the quantitative prefactor assumes the overdamped TDGL dynamics derived for the dirty, near-$\Tc$ regime $\Delta_{0}(T)\ll T$; away from this window the attempt frequency is uncertain (Appendix~\ref{app:prefactor}), although the barrier asymmetry, being a property of the Gibbs distribution, is insensitive to this uncertainty. We also assumed a wire that is one-dimensional for the order parameter, with transverse dimensions small compared with $\xi$, while the electron motion in it is diffusive and two-dimensional, and neglected orbital pair breaking of the in-plane field, appropriate for film thickness small compared with the magnetic length.

\section{Summary and outlook}
\label{sec:discussion}

We have generalized the LAMH theory of thermally activated phase slips to superconducting wires with broken inversion symmetry, using the microscopically derived generalized TDGL theory of a disordered Rashba superconductor with an in-plane field. The main results are as follows. The phase-slip activation barrier acquires a correction odd in both current and field, Eq.~\eqref{eq:dFbar}, equivalent to an anomalous phase offset of the phase-slip saddle; the wire is, at the level of its resistive fluctuations, a distributed counterpart of the $\varphi_{0}$ junction. The leading Lifshitz invariant cancels exactly from the fixed-current barrier, and the effect is controlled by the subleading invariants through the same effective cubic coefficient as in the diode problem, with the parameter-free relation $\chi=\sqrt6\,\eta$ between the anomalous phase and the diode efficiency, verified independently by matching the barrier to the diode-shifted critical currents at depairing. The two sectors of the theory play distinct roles. The thermodynamic invariants determine the activation exponent, whereas the kinetic invariant, which is tied to the noise by the fluctuation-dissipation theorem, leaves the Gibbs distribution unchanged and enters only the prefactor, at relative order $k_{B}T/\DF_{0}$, except near a sign change of $a_{3}$, where it becomes the leading source of nonreciprocity. The observable consequence is a magnetochiral anisotropy of the wire resistance that is enhanced relative to the diode asymmetry by a factor of order the barrier height over temperature and appears as a field-antisymmetric splitting of the exponential slopes of differential-resistance traces, Eq.~\eqref{eq:dvdi}, with a distinctive in-plane angular dependence and a built-in consistency test against the diode efficiency of the same device.

We now compare this work with earlier extensions of phase-slip theory beyond the standard LAMH framework. Ref. \cite{Zharov2007} constructed the thermal phase-slip saddle point of a clean single-channel wire from the Eilenberger equations and obtained the activation barrier at all temperatures below $\Tc$. Their result concerns the static barrier: in a clean, momentum-conserving wire the barrier is well defined, but converting a barrier into a dc resistance requires a momentum-relaxing bath and a dissipative order-parameter dynamics, which the clean limit does not supply. Our problem is complementary: the disordered wire furnishes the momentum sink and the Keldysh-derived TDGL furnishes internally consistent dissipation and noise, at the price of being confined to the vicinity of $\Tc$; and our focus is the nonreciprocal physics absent from their inversion-symmetric model. Ref. \cite{PesinAndreev2006} identified fluctuations of a different nature: nonperturbative saddle points of the replica sigma model of a disordered wire, involving the diffusive electron modes and the Coulomb field rather than the order parameter alone, invisible to any GL functional and yielding a negative, field-even magnetoresistance. Both mechanisms coexist with the one studied here. In the centrosymmetric models considered in these works both are reciprocal, so that, like the conventional LAMH background, they do not contribute to the field-antisymmetric part of the resistance isolated in Fig.~\ref{fig:dvdi}(c). Their nonreciprocal generalizations, discussed below, would contribute to the same symmetry channel as the effect studied here.

We mention several extensions. First, experiments observe thermally activated behavior well below the GL window, and the clean-wire results of Ref.~\cite{Zharov2007} together with the quasiclassical treatment of Rashba and exchange fields pioneered for $\varphi_{0}$ junctions \cite{Buzdin2008,BergeretTokatly2014,Konschelle2015} suggest that the nonreciprocal barrier can be computed at the Eilenberger level at all temperatures. In the helical-band basis the spectral parameter is shifted by the deviation of the condensate momentum from the helical minimum, and the kink-type saddle is expected to persist with computable odd corrections. Deriving the corresponding low-temperature anomalous phase, and the appropriate dissipative kinetics beyond TDGL, is an open problem. Second, the present TDGL framework, being built on the perturbative cooperon expansion, does not capture nonperturbative configurations of the sigma-model manifold of the type underlying the saddle points of Refs.  \cite{PesinAndreev2006,Kamenev2000}; a nonreciprocal generalization of that physics, with spin-orbit coupling and field built into the sigma-model action, may produce contributions odd in both current and field and remains unexplored. Third, at lower temperatures quantum phase slips take over. The instanton action should acquire corrections odd in the field, which suggests a quantum diode effect in the phase-slip resistance that has not yet been studied. Unlike the thermal case, the kinetic invariant is expected to enter the instanton action itself, since the quantum exponent contains the dissipative dynamics. Fourth, the magnetochiral response vanishes for a field along the wire, which is the geometry used to reach the topological regime of proximitized Rashba nanowires \cite{Lutchyn2010,Oreg2010}. This zero is exact: for a field parallel to the current, reflection in the plane perpendicular to the wire preserves both the polar axis and the field while reversing the current. It is therefore a symmetry of the Rashba geometry and requires $V(I,B)=-V(-I,B)$ to all orders, whether the wire is topological or not. A longitudinal response requires lower symmetry, such as Dresselhaus spin-orbit coupling or a chiral crystal structure, for which the Lifshitz vector acquires a component along the field. In that case the present theory applies with $\sin\theta_{B}$ replaced by the projection of $\hat n$ on the wire axis. In the topological regime itself the GL description used here does not apply, and phase slips are coupled to the fermion parity of the wire \cite{Fidkowski2011}. Whether this produces a distinct nonreciprocal signature in geometries of reduced symmetry is an open question. Finally, away from equilibrium the fluctuation-dissipation relation that protects the Gibbs distribution no longer holds, and friction and noise can be nonreciprocal independently \cite{Liu2026}. Phase-slip noise in noncentrosymmetric wires is therefore of interest for both theory and experiment.

\begin{acknowledgments}
The work was supported by NSF Grant No. DMR-2452658 and H. I. Romnes Faculty Fellowship provided by the University of Wisconsin-Madison Office of the Vice Chancellor for Research and Graduate Education with funding from the Wisconsin Alumni Research Foundation. The author acknowledges the use of the large language model Claude (Anthropic, model Fable 5.2) in the preparation of the manuscript, including, in particular, numerical calculations, graphics, and symbolic verification of analytical results. The author conceived and directed the project, verified the results, and takes full responsibility for the content of the paper.
\end{acknowledgments}

\appendix

\section{Closed-form geometry of the phase-slip saddle}
\label{app:saddle}

All saddle integrals follow from the substitution $t=\tanh(\beta x)$, under which $g\equiv f^{2}=2k^{2}+(1-3k^{2})t^{2}$ and $\dd x=\dd t/[\beta(1-t^{2})]$, with $\beta=\sqrt{(1-3k^{2})/2}$ and $j=k(1-k^{2})$:
\begin{align}
\int\big(g-g_{\infty}\big)\dd x&=-2\sqrt2\sqrt{1-3k^{2}},
\label{eq:app-deficit}\\
\delta\phi(k)=j\!\int\!\Big(\frac1g-\frac1{g_{\infty}}\Big)\dd x
&=2\arctan\frac{\sqrt{2(1-3k^{2})}}{2k},
\label{eq:app-dphi}\\
\Delta F(k)\big/E_{\xi}
&=\frac{4\sqrt2}{3}\sqrt{1-3k^{2}} .
\label{eq:app-DF}
\end{align}
The phase integral has the closed form quoted in Eq.~\eqref{eq:phase}; the coefficient of the arctangent equals unity identically, because $j\sqrt{\Delta g}/(\beta g_{\infty}\sqrt{g_{0}})=1$ with $g_{0}=2k^{2}$, $\Delta g=1-3k^{2}$, $g_{\infty}=1-k^{2}$, and its asymptotics reproduce Eq.~\eqref{eq:app-dphi}. The Gibbs barrier at fixed current,
\begin{align}
\frac{\Delta G_{+}(k)}{E_{\xi}}
&=\frac{4\sqrt2}{3}\sqrt{1-3k^{2}}
-4j\arctan\frac{\sqrt{2(1-3k^{2})}}{2k},
\label{eq:app-DG}
\end{align}
vanishes at depairing as $\Delta G_{+}=\frac{48\cdot3^{1/4}}{5}(k_{c}-k)^{5/2}E_{\xi}$, equivalent to the standard $(1-I/I_{c})^{5/4}$ law. All formulas of this appendix were verified symbolically and numerically.

The on-shell identity \eqref{eq:cubicdensity} follows directly: with $u=f\e^{\ii\phi}$, $\phi'=j/f^{2}$, and $\phi''=-2jf'/f^{3}$,
\begin{align}
\Imop(u'^{*}u'')&=2f'^{2}\phi'+ff'\phi''-ff''\phi'+f^{2}\phi'^{3}
\nonumber\\
&=-j\,\frac{f''}{f}+\frac{j^{3}}{f^{4}}
=j\,(1-f^{2}),
\end{align}
where the first two terms cancel identically and the last equality uses the saddle equation $f''/f=-1+f^{2}+j^{2}/f^{4}$.

Finally we record the depairing consistency check of Sec.~\ref{sec:diode-link}. Near $k_{c}$ one has $\sqrt{1-3k^{2}}\simeq(2\sqrt3)^{1/2}(k_{c}-k)^{1/2}$ and $k_{c}-k\simeq[(j_{c}-j)/\sqrt3]^{1/2}$, so the perturbation \eqref{eq:dFbar} behaves as
$\delta\Delta\bar F=4\sqrt2\cdot3^{-11/8}\chi\,(j_{c}-j)^{1/4}E_{\xi}$,
while shifting the critical current in the unperturbed barrier, $\Delta G_{+}(j;j_{c}\to j_{c}(1+\eta))$, produces
$24\cdot3^{-15/8}\eta\,(j_{c}-j)^{1/4}E_{\xi}$.
The two coincide if and only if $\chi=\sqrt6\,\eta$, which is Eq.~\eqref{eq:chi} derived independently from the small-current integrals. At small currents the analogous ratio of the exact asymmetry to the shifted-critical-current estimate is $\sqrt6\sqrt{1-3k^{2}}/\delta\phi(k)\to\sqrt6/\pi\simeq0.78$.

\section{Gradient-ordering ambiguities}
\label{app:ordering}

The coefficients $\alpha_{4}$ and $\mathfrak{b}_{2}$ are extracted from plane-wave data: $a(q)$ and $b(q)$ are, by construction, the energy densities of single-harmonic configurations. The correspondence between a polynomial in $q$ and a local Hermitian operator is not unique, and since the phase-slip saddle is strongly nonuniform, the choice might appear to matter. As an example, consider the symbol $q^{2}|\Delta|^{4}$ of the $\mathfrak{b}_{2}$ term. Two admissible realizations are
\begin{equation}
\mathcal{Q}_{1}=|\Delta|^{2}|\partial_{x}\Delta|^{2},
\qquad
\mathcal{Q}_{2}=-|\Delta|^{2}\,\Reop(\Delta^{*}\partial_{x}^{2}\Delta).
\label{eq:app-Q12}
\end{equation}
Writing $\Delta=f\e^{\ii\theta}$ one finds $\mathcal{Q}_{1}=f^{2}(f'^{2}+f^{2}\theta'^{2})$ while $\mathcal{Q}_{2}=f^{2}(f^{2}\theta'^{2}-ff'')$: the two coincide on any plane wave, where $f'=0$, but differ on the saddle by the amplitude-gradient density $f^{2}(f'^{2}+ff'')$, which is not a total derivative once weighted by $f^{2}$. Since the phase-slip asymmetry rests on near-cancellations among the couplings, this ambiguity must be addressed. Three observations settle the issue.

First, the ambiguous pieces are even-sector densities at the $\alpha_{4}$, $\mathfrak{b}_{2}$ gradient order. In the saddle geometry every term of this order corrects the barrier only at relative order $|\epsB|$, and all of them, ambiguous parts included, are dropped consistently with the accuracy of the GL expansion itself.

Second, the only part of these even operators that survives in the odd sector is the cross term linear in $q_{0}$ generated by the helical shift $\Delta=\e^{\ii q_{0}x}\psi$, and that part is realization independent. For the pair \eqref{eq:app-Q12}, substituting $\partial_{x}\to\partial_{x}+\ii q_{0}$ and expanding to linear order in $q_{0}$ gives
\begin{equation}
\mathcal{Q}_{1},\;\mathcal{Q}_{2}\;\to\;2q_{0}\,|\psi|^{2}\,\Imop(\psi^{*}\partial_{x}\psi),
\label{eq:app-Qshift}
\end{equation}
identically for both realizations: the ambiguity cancels in the cross term. For the $\alpha_{4}$ symbol $q^{4}|\Delta|^{2}$ the realizations $|\partial_{x}^{2}\Delta|^{2}$ and $\Reop(\Delta^{*}\partial_{x}^{4}\Delta)$ give the linear parts $4q_{0}\,\Imop(\psi'^{*}\psi'')$ and $-4q_{0}\,\Imop(\psi^{*}\psi''')$, which differ by the total derivative $\frac{\dd}{\dd x}\Imop(\psi^{*}\psi'')$; the corresponding boundary term cancels in the barrier difference because the saddle approaches the uniform state at both ends, where $\Imop(\psi^{*}\psi'')$ takes equal values. The general statement follows the same pattern: two realizations of one symbol differ by a density that vanishes on all plane waves, and its $q_{0}$-linear part is again of this type; within the polynomial expansion at the retained order, the basis of odd densities modulo total derivatives is exhausted by the operators $\mathcal{O}_{3}$ and $\mathcal{O}_{1}$ of Eq.~\eqref{eq:operators}, so the plane-wave symbols fix the odd couplings uniquely. Remaining odd densities that vanish on plane waves belong to higher orders of the gradient and amplitude expansion, beyond the accuracy of the microscopic input, and are suppressed by a further factor of $|\epsB|$ in the barrier.

Third, the first-order barrier shift is $\delta F_{\rm odd}$ evaluated on the unperturbed saddle, where the on-shell identities \eqref{eq:quarticdensity} and \eqref{eq:cubicdensity} reduce both retained operators to the single integral \eqref{eq:deficit}, and any total-derivative term cancels between the saddle and the metastable state. The combination $\chi=\sqrt6\,\eta$ is therefore independent of the operator-ordering convention.

\section{Fluctuation determinant and the prefactor}
\label{app:prefactor}

Linearizing around the zero-current saddle $u_{s}=\tanh(x/\sqrt2)$, the fluctuation operators for the real and imaginary parts of $u$ are the P\"oschl--Teller Hamiltonians
\begin{align}
H_{+}&=-\partial_{x}^{2}+2-3\,{\rm sech}^{2}(x/\sqrt2),
\nonumber\\
H_{-}&=-\partial_{x}^{2}-{\rm sech}^{2}(x/\sqrt2),
\label{eq:app-ops}
\end{align}
with exact spectra: $H_{+}$ has bound states at $0$ (translation mode $\propto{\rm sech}^{2}$) and $3/2$, continuum from $2$; $H_{-}$ has the negative mode $-1/2$ ($\propto{\rm sech}$, the unstable direction that unwinds the phase), the global phase mode at $0$ ($\propto u_{s}$), and a gapless continuum. The Kramers--Langer rate for the overdamped dynamics \eqref{eq:TDGL} requires the ultraviolet-finite determinant ratios with zero modes removed, which we evaluated by numerical diagonalization:
\begin{equation}
\frac{\det H_{+}^{0}}{\det{}'H_{+}}=24,
\qquad
\frac{\det{}'H_{-}^{0}}{\big|\det{}'H_{-}\big|}\to1 .
\label{eq:app-dets}
\end{equation}
Here the superscript $0$ denotes the uniform-state operators, and the prime indicates that the zero mode is omitted: the translation mode of $H_{+}$ and the global phase mode of $H_{-}$ and $H_{-}^{0}$. The negative eigenvalue of $H_{-}$ enters through its absolute value. The arrow denotes the limit $L\gg\xi$: since the spectrum of $H_{-}$ is gapless, the second ratio approaches unity with a correction of relative order $\xi/L$.
The first ratio also follows in closed form from the Gel'fand--Yaglom formula, $\det{}'H_{+}/\det H_{+}^{0}=\int\psi_{0}^{2}\,\dd x/2mA^{2}=1/24$, where $\psi_{0}={\rm sech}^{2}(x/\sqrt2)\to A\,\e^{-m|x|}$ is the zero mode, with $A=4$ and $m=\sqrt2$. Assembling Langer's formula with the translation-mode volume $(L/\xi)\big(E_{\xi}\!\int u_{s}'^{2}\dd x/\pi k_{B}T\big)^{1/2}=(L/\xi)(\DF_{0}/2\pi k_{B}T)^{1/2}$, the unstable growth rate $\lambda_{+}=1/2\tau_{\rm GL}$, and the correspondence between the modes of the saddle and of the uniform state yields Eq.~\eqref{eq:MH} with the closed-form constant $\sqrt3/2\pi^{3/2}\simeq0.155$, coinciding with the McCumber--Halperin result \cite{McCumber1970}.

Ref. \cite{Golubev2008} criticized the TDGL basis of this prefactor and, evaluating the imaginary part of the free energy within a microscopic effective action, obtained a rate larger by the factor of order $T_{c}\tau_{\rm GL}\sim(1-T/\Tc)^{-1}$, with the near-depairing current dependence $(1-I/I_{c})^{5/8}$ in place of $(1-I/I_{c})^{15/8}$. A direct comparison shows that the static parts of the two calculations coincide: the operators of their Appendix are Eqs.~\eqref{eq:app-ops} in rescaled units, their phase-sector determinant equals our second ratio in Eq.~\eqref{eq:app-dets}, and their amplitude-sector factor $\tfrac{2\sqrt6}{\sqrt\pi}(L/\xi_{\rm GZ})\sqrt{\delta F/T}$ coincides, after the length conversion $\xi_{\rm GZ}=\sqrt2\,\xi$, with our $\sqrt{12/\pi}\,(L/\xi)\sqrt{\DF_{0}/T}$ built from Eq.~\eqref{eq:app-dets}. The disagreement is therefore confined to the dynamical conversion factor: $\lambda_{+}/2\pi=1/4\pi\tau_{\rm GL}$ here versus a crossover temperature of order $T_{c}$ there, the latter estimated by matching the quantum-phase-slip action to the activation exponent. In the classical regime the imaginary-free-energy method requires the conversion factor to be the real-time growth rate of the unstable mode under the actual dissipative dynamics; for the microscopically derived overdamped kernel \eqref{eq:kinetic}, valid at real frequencies $\omega\ll T$ in the dirty regime $\Delta_{0}(T)\ll T$, that rate is $1/2\tau_{\rm GL}$, and Eq.~\eqref{eq:MH} follows. The matching estimate instead uses the temperature at which quantum decay becomes competitive, which is a different quantity when the thermal saddle and the quantum instanton are characterized by parametrically different frequencies. We therefore adopt Eq.~\eqref{eq:MH} for the regime of this paper, while noting that the controversy is unresolved experimentally, since the prefactor enters $\ln R$ only additively. None of the nonreciprocal observables of Sec.~\ref{sec:results} depend on the choice: the attempt factor is even in $(I,B)$ at leading order in either scheme, and the static determinants are Gibbsian and independent of the kinetic invariant exactly.

\section{Microscopic couplings across the spin-relaxation crossover}
\label{app:coeffs}

All couplings of Eqs.~\eqref{eq:F}--\eqref{eq:kinetic} were derived in Ref.~\cite{Liu2026} from a single kernel controlled by $g=\Gt/4\pi T$, $\Gt=2(\aR\pF)^{2}\tau$; every odd coupling points along the axis $\hat n \parallel \hat n_{R}\times\bm h$ and carries one factor of $\aR h$. Table~\ref{tab:couplings} lists the two limits. The even coefficients
\begin{equation}
\alpha_{4}=\frac{28\zeta(3)}{\pi^{4}}\,\xiGL^{4},
\qquad
\mathfrak{b}_{2}=\frac{4\pi^{2}}{21\zeta(3)}\,\xiGL^{2}
\label{eq:app-even}
\end{equation}
are field independent. The quartic-sector entries are the ladder-complete vertex values, fixed by the diagrammatic evaluation of Ref.~\cite{Hasan2025}, larger than the single-mode estimates by $8/3$ ($\mathfrak b_1$) and $4/3$ ($\mathfrak b_{2}$); with the single-mode vertex the bracket of Eq.~\eqref{eq:bracket} changes sign, so this distinction is qualitative. The signs of $\Lambda$, $\Xi$, and $\rho$ do not change across the crossover \cite{Liu2026}. The cubic channel, however, enters $\chi$ through $\Xi_{\rm eff}=\Xi-4\alpha_{4}q_{0}$, which is positive in the weak-DP limit and vanishes in the DP-dominated limit, where $\Xi=4\alpha_{4}q_{0}$ follows from Table~\ref{tab:couplings} and Eq.~\eqref{eq:app-even}. In that limit $\chi$, like $\eta$, is determined by $\mathfrak{b}_{\rm eff}$ alone, and any disorder-induced reversal of $\chi$ originates in the quartic sector \cite{Liu2026}.

\begin{table}[b]
\caption{Nonreciprocal couplings in the two limits of the Dyakonov--Perel crossover \cite{Liu2026}. All entries are quoted for the geometry of Sec.~\ref{sec:geometry}.}
\begin{ruledtabular}
\begin{tabular}{lcc}
Coupling & $\Gt\ll4\pi T$ & $\Gt\gg4\pi T$\\
\colrule
$\Lambda$ & $\dfrac{7\zeta(3)}{2\pi^{2}}\dfrac{(\aR\pF\tau)^{2}}{T^{2}}\aR h$ & $\dfrac{\pi\tau}{2T}\,\aR h$\\[8pt]
$\Xi$ & $\dfrac{\pi D}{16}\dfrac{(\aR\pF\tau)^{2}\aR h}{T^{3}}$ & $\dfrac{7\zeta(3)}{2\pi^{2}}\dfrac{D\aR h\tau}{T^{2}}$\\[8pt]
$\rho$ & $\dfrac{(\aR\pF\tau)^{2}}{2T^{2}}\,\aR h$ & $\dfrac{28\zeta(3)}{\pi^{3}}\dfrac{\aR h\tau}{T}$\\[8pt]
$\mathfrak b_1$ & $\dfrac{124\zeta(5)}{7\zeta(3)}\dfrac{(\aR\pF\tau)^{2}}{\pi^{2}T^{2}}\aR h$ & vertex crossover \cite{Liu2026}\\[6pt]
$q_{0}$ & $\dfrac{14\zeta(3)}{\pi^{2}}\,g\,\dfrac{4\aR h}{\vF^{2}}$ & $\dfrac{4\aR h}{\vF^{2}}$\\
\end{tabular}
\end{ruledtabular}
\label{tab:couplings}
\end{table}

\bibliography{refs}

\end{document}